# Cytoskeletal network morphology regulates intracellular transport dynamics

**Authors:** David Ando[1], Nickolay Korabel[1,2], Kerwyn Casey Huang[3,4,*], Ajay Gopinathan[1,*]

**Affiliations:**

[1]Department of Physics, University of California at Merced, Merced, CA, USA

[2]School of Mathematics, University of Manchester, Manchester, UK

[3]Department of Bioengineering, Stanford University, Stanford, CA, USA

[4]Department of Microbiology and Immunology, Stanford University School of Medicine, Stanford, CA, USA

*Correspondence to: kchuang@stanford.edu, agopinathan@ucmerced.edu

# ABSTRACT

Intracellular transport is essential for maintaining proper cellular function in most eukaryotic cells, with perturbations in active transport resulting in several types of disease. Efficient delivery of critical cargos to specific locations is accomplished through a combination of passive diffusion and active transport by molecular motors that ballistically move along a network of cytoskeletal filaments. Although motor-based transport is known to be necessary to overcome cytoplasmic crowding and the limited range of diffusion within reasonable time scales, the topological features of the cytoskeletal network that regulate transport efficiency and robustness have not been established. Using a continuum diffusion model, we observed that the time required for cellular transport was minimized when the network was localized near the nucleus. In simulations that explicitly incorporated network spatial architectures, total filament mass was the primary driver of network transit times. However, filament 'traps' that redirect cargo back to the nucleus caused large variations in network transport. Filament polarity was more important than filament orientation in reducing average transit times, and transport properties were optimized in networks with intermediate motor on and off rates. Our results provide important insights into the functional constraints on intracellular transport under which cells have evolved cytoskeletal structures, and have potential applications for enhancing reactions in biomimetic systems through rational transport network design.

# INTRODUCTION

The transport of intracellular cargos occurs in virtually all eukaryotic cells and is essential for many cellular functions. Whereas diffusion is sufficient for the transport of small (~nm) molecules, diffusion becomes prohibitively slow for large cargos traversing extremely crowded and relatively large cellular distances; these cargos thus require an active transport system (1, 2). Such intracellular transport is carried out by molecular motors that move along a complex system of cytoskeletal polymers such as actin filaments, actin bundles, and microtubules (3). The three classes of molecular motors (myosin, kinesin, dynein) that transport cargo along the cytoskeleton are broadly categorized by their transport behavior and by the filament type upon which they move. Most myosins travel toward the barbed (+) end of actin filaments (4), which tend to form random networks (5). Most kinesins move toward the plus end of microtubules, which are typically radially oriented and polarized with the (+) end pointed away from the nucleus (toward the cell membrane) (6), while dyneins transport cargo toward the minus end (-) of their microtubule track (7). Despite a wealth of information about the biophysical properties of motors (8) and the filaments upon which they walk (9), we have little knowledge of the relative importance of various transport variables in the vast phase space of network, motor, and cargo properties, especially in the context of explicit models of physical network architecture. Computational models provide the means to identify how these transport variables influence outcomes such as the mean and variation in transport time between cellular locations that may impact cellular fitness.

Conceptual frameworks of active intracellular transport (6, 10) have considered a cytoskeletal network with organized microtubules that serve as tracks for long-distance transport, while random actin-filament networks are utilized for local transport. Microtubules are polarized and radially oriented away from the nucleus and the centrosome (9), enabling transport from the nucleus to the periphery or vice versa in an approximately linear manner. In conjunction, actin filaments are thought to underlie local transport from the microtubule network to the remainder of the cell (11) via a randomly oriented network that is distributed throughout the cytoplasm. While there have been careful studies of motor properties (8) and of the structure of individual network filaments (9), we still lack understanding of how motor properties coordinate with the topology of cytoskeletal networks to yield efficient, reliable transport.

Previous computational studies of intracellular transport mainly relied on reaction-hyperbolic partial differential equation modeling and so-called 'virtual network' models that are based on the theory of intermittent dynamics in the context of search processes (2, 12-15). Virtual network models assume that cargos repeatedly bind and unbind motors (5, 15), and as a result exhibit phases of diffusion for random lengths of time that are interrupted by ballistic excursions of constant velocity and direction. Because the ballistic motion representing motor binding is assumed to be uniform in space and direction, heterogeneity in the network architecture is usually ignored. A previous study used an explicit filament network model to determine the effect of motor processivity on intracellular transport times (16). However, it has not yet been determined how the morphological properties of the network itself dictate transport outcomes relative to motor properties.

Here, we use simulations to determine how cells regulate transport via changes in network and motor properties, such as filament localization, polarity, and orientation and motor mobility and binding. We first use a simple continuum model of increased bulk diffusion in a shell representing the cytoskeletal network to show that transport time from the nucleus to the periphery is minimized when the shell is localized near the nucleus, demonstrating the potential importance of the overall spatial configuration of the network. Simulations of ballistic motion along polarized filaments with explicit spatial extent reveal that particular filament arrangements near the nucleus can constitute 'traps' (regions in which cargos spend increased amounts of time) that lead to large variations in the time required to achieve transport from the nucleus to the cell membrane. This variability, which can be mitigated by distributing the network mass over more filaments with shorter lengths, may be an important constraint on network organization. We also demonstrate that polarizing filaments with the direction of transport facing away from the nucleus was significant in reducing transit time, although the precise distribution of the angles of orientation has little effect. Finally, we determine that high on rates and/or low off rates, potentially representative of the binding properties of cargo complexes with multiple motors, actually slow cargo transport and increase variability, suggesting that single motors can more robustly transport cargo than a collection of multiple motors over random networks.

## MATERIALS AND METHODS

For our continuum model diffusion simulations, cargo complexes diffused by taking random steps of fixed length (100 nm), with the time required per step determined by the bulk diffusion constant of the region in which the cargo was located. For explicit networks, simulations were initialized by placing circular cargo-motor complexes 100 nm in radius at random locations on the nuclear boundary. These complexes subsequently diffused randomly in steps of fixed length (100 nm) until they bound a filament with which the cargo overlapped at rate $k_{on}$. Filaments were represented as line segments with polarity. During network initialization, the centers of mass of filaments with a portion lying outside the cell or inside the nucleus were shifted inward or outward, respectively, along the radial direction until the filament was completely inside the cell, thus ensuring that all filaments were of equal length. Active motor transport occurred in 100-nm steps, with displacement in the direction of the filament's (+) end at a velocity of 1 µm/s. During active transport, cargos dissociated at the motor off rate $k_{off}$ or when they encountered the end of the filament. After unbinding, cargos were moved precisely adjacent to but not overlapping the filament. An effective cytoplasmic viscosity for cargos of 0.05 Pa s (1, 17, 18) was assumed, resulting in a diffusion constant of $D = 0.051$ µm$^2$/s for a cargo-motor complex with radius 100 nm. Additional methods details appear in the Supporting Material.

RESULTS

**Localizing transport systems near the nucleus minimizes cargo-transport time**

The endoplasmic reticulum, an organelle that is located near and is continuous with the outer nuclear membrane, plays a central role in membrane and protein biosynthesis (19). We therefore focused our modeling on the transport of cargos that originate from the surface of an impenetrable nuclear membrane and terminate at the cell (cytoplasmic) membrane. Since imaging studies generally visualize cells adhered to a substrate, which tends to produce a flattened morphology, we focused our modeling on transport in two dimensions. Although eukaryotic cell shape and size can vary across a broad range, we simulated a circular, two-dimensional cell with radius $R$ = 10 μm, the typical size of well-studied eukaryotic cell lines (20).

To investigate whether the spatial localization of the cytoskeletal network affects the time required to transport cargo from the nucleus to the membrane, we initially implemented a simple continuum diffusion model of cargos that originate at the precise center of a cell (in the absence of a nucleus) and are transported solely by diffusion with constant $D$; transport terminates when the cargo reaches the cell membrane (Fig. 1A). The equation describing the concentration of cargos $c(x, y, t)$ diffusing in two dimensions is the diffusion equation:

$$\frac{\partial c}{\partial t} = D \left( \frac{\partial^2 c}{\partial x^2} + \frac{\partial^2 c}{\partial y^2} \right). \tag{1}$$

For the transport problems we consider, we assume a constant source at either the center of the cell or at the nuclear boundary, and zero cargo concentration along the cell boundary.

At steady state, Eq. (1) transformed into polar coordinates becomes

$$D \left( \frac{\partial^2 c}{\partial x^2} + \frac{\partial^2 c}{\partial y^2} \right) = \left( \frac{\partial^2 c}{\partial r^2} + \frac{1}{r} \frac{\partial c}{\partial r} + \frac{1}{r^2} \frac{\partial^2 c}{\partial \theta^2} \right) = 0. \qquad (2)$$

Since the system is symmetric with respect to $\theta$, the solution is

$$c(r, \theta) = C_1 + C_2 \log r. \qquad (3)$$

The spatial distribution of cargo residence time can then be calculated by applying the appropriate boundary conditions; see Supporting Material for full derivation and analytical solutions.

In this simple scenario, cargo residence time as a function of distance from the cell center peaked at an intermediate radius (Fig. 1B). To introduce the effects of active transport in a simplified manner, we represented a cytoskeletal network as an annulus of fixed width $w$ whose inner radius is $R_a$, within which the diffusion constant is increased to $D_a = 100D$ to account for rapid motor transport along a random network. This annulus was placed outside a nucleus with radius $R_n$. For small nuclear radii ($R_n \lesssim 2$ μm) and an annulus width $w = 3$ μm, the mean first passage time (MFPT) of cargo transport from the nuclear surface to the cell membrane was minimized when the inner boundary of the annulus was placed away from the nucleus, with the center of the annulus at a radius of ~3.5 μm (Fig. 1C). This minimal MFPT position corresponds approximately to the location of the peak residence time in the absence of a network (Fig. 1B), suggesting that localization of enhanced diffusion around regions of peak residence times provides the greatest reduction in MFPT. Annuli closer to the cell membrane led to higher MFPTs (Fig. 1C), even though the total annulus area increased linearly with $R_a$. When we instead fixed the area of the annulus such that $w$ was inversely proportional to $R_a$, the MFPT increased monotonically as the annulus was shifted away from the nucleus, regardless of $R_n$ (Fig. S1). Thus, for cells with $R_n \gtrsim R/4$, a range encompassing typical nuclear to membrane size ratios, accelerating diffusion closer to the nucleus generally reduces MFPT.

**Filament network architecture affects transport-time variability**

To determine whether specific network architectures affect transport properties, we moved away from continuum diffusion simulations and introduced active transport via ballistic motion of cargo-motor complexes along explicitly modeled filaments. In our model, as *in vivo*, the filaments are characterized by their length, orientation, and polarity. Orientation and polarity of the filaments are defined relative to the cellular geometry. The filament angle $\theta$ is defined as the angle between the filament's direction vector (from the filament midpoint to its plus end) and a vector pointing radially outward. $\theta$ modulo $\pi$ radians is defined as the orientation of a filament. Filament polarity is defined by the sign of the projection of the filament's direction vector on the vector pointing radially outward, which is the sign of $\cos(\theta)$. Thus, a positive or negative value corresponds to outward or inward polarity, respectively. For example, for a network in which all filaments have an orientation of $\theta = 0$ or $\pi$, all filaments would be radially oriented with their polarities defined by whether the (+) or (-) end is closest to the cell membrane. By contrast, a network could have outwardly polarized filaments with random orientations, which would correspond to values of $\theta$ uniformly distributed between $-\pi/2$ and $+\pi/2$. Motor-cargo complexes bind a filament with a specified on rate $k_{on}$, walk ballistically in the direction determined by the filament polarity, and unbind from the filament with an off rate $k_{off}$ or fall off the end of the filament. The cargo then returns to the cytoplasm and diffuses with constant $D$. *In vivo*, many individual molecular motors such as kinesin, myosin-V, and dynein have processivities of ~1-2 μm during transport (8, 18, 21-23); we start by assuming a mean processivity of 1 μm.

We first focused on randomly oriented and positioned filament networks (Methods; Fig. 2A), which can approximate disordered actin-filament or microtubule networks as observed *in vivo* (5, 24). We assumed a nuclear radius of 4.8 μm, which is approximately consistent with the observed ratio between nuclear and membrane size (25, 26). The filaments were located in the outer 5 μm of a two-dimensional cell 10 μm in radius. The 0.2-μm buffer between the nuclear membrane and the region occupied by filaments allowed cargos to reach the ends of filaments near the nuclear membrane without interference from the nuclear surface, which was treated as a reflecting boundary condition. Up to 500 filaments were simulated, which is equivalent to ~10-fold the microtubule density in monkey kidney epithelia cells (27). The average microtubule length in human foreskin fibroblast cells is 5.45 μm (28), which is roughly comparable to our maximum simulated filament length of 5 μm. Similarly, the range of parameters simulated for filament number and filament length are representative of actin filaments in the cytoplasmic bulk, which have lengths of ~1-3 μm (29, 30). The density of actin filaments in flattened BSC-1 epithelial cells can be estimated via STORM imaging to be ~3-10 μm of filament per μm$^2$ (31), which is roughly comparable to 140 - 470 5-μm filaments present in our simulation geometry.

In our model, a motor-cargo complex was represented by a circle with radius 100 nm, which is roughly consistent with the typical size range of a molecular motor (30-50 nm (32)) bound to a vesicle (20-600 nm (33)). Given that cells are ≳200 nm in height (34), we assumed that these particles experience no excluded-volume interactions with the network filaments, enabling our model to more effectively represent the three-dimensional nature of living cells. For simplicity, there were no interactions between cargo-motor complexes in our simulations. When a cargo spatially overlapped with a filament, binding within a time step *dt* occurred with probability $k_{on}dt$. Binding was followed by ballistic movement at a constant velocity in the direction indicated by the filament's polarity, until the cargo either dissociated from its track or reached the end of the filament.

We first determined the MFPT of cargo-motor complexes moving on random networks as a function of the primary network-topology parameters of filament length and number (Fig. 2B). For each combination of filament length and number, we simulated 3,000 cargos on each of 10,000 network realizations. We used physiologically relevant parameters motivated by *in vitro* experiments on kinesin/myosin (21) (Methods), with motor velocity $v = 1$ μm/s, bulk diffusion $D = 0.051$ μm$^2$/s, $k_{on} = 5$/s, and $k_{off} = 1$/s. The MFPT averaged over network realizations and cargos was approximately constant over a wide range of filament lengths and numbers in networks with fixed total filament mass (Fig. 2B). However, the standard deviation of MFPT over network realizations was high relative to the mean along curves of constant mass in networks with fewer filaments (Fig. 2C). Thus, specific topological features of a random transport network can dramatically affect transport rates such that architectures with a higher number of shorter filaments have lower variation in MFPT.

**Network-transport variability results from rare cargo traps**

Over many network realizations the distribution of MFPTs was much broader for networks with fewer filaments than for networks with many filaments (Fig. 2C,3A). To determine the origin of the large standard deviation in MFPT in networks with relatively few longer filaments, we visually inspected networks that had particularly large MFPTs. We spatially binned the residence time averaged over cargos across the cytoplasm and noticed that for a network with large MFPT (220 s) relative to the average MFPT (82 s), there was a hotspot of >10-fold higher residence times near the nucleus in a region with multiple filaments pointed inward that acted as a sink for cargo (Fig. 3B). To verify that this architecture was directly responsible for the large range of MFPTs, we reversed the polarity of each filament in the network and observed a drop in the MFPT to 62 s, without any residence-time hotspots (Fig. 3C). We binned space into 0.04-$\mu m^2$ regions and defined traps using a residence-time threshold of 10 s/$\mu m^2$. This choice of residence time threshold guaranteed that there were no spurious traps in the absence of filaments over a wide range of spatial bin sizes (Fig. S7). We scanned 1,000 network realizations and found that the total area of trap regions was strongly related to the MFPT of a network (Fig. 3D), further supporting the hypothesis that traps are a major underlying cause of increases in MFPT for a given number of filaments of fixed length. Since a reasonable fraction (79%) of network realizations do not have traps, reversing filament polarity within a given network architecture generally removed traps from networks with large trap area, reducing MFPT (Fig. 3E).

With a 0.2-$\mu m$ buffer surrounding the nucleus, traps generally occurred near the nucleus, with 86% of the total trap area in traps with a centroid within 1 $\mu m$ of the nuclear boundary (Fig. S2A). We sought to understand the origin of this localization by increasing the size of the buffer region around the nucleus in our simulations from 0.2 to 1.0 $\mu m$, which we accomplished by shrinking the nuclear radius by 0.8 μm. To make MFPTs comparable across simulations with small and large buffers, we ignored cargo-residence times in the extended buffer region. In simulations with the larger buffer, both overall trap number and MFPT variability over network realizations decreased strongly (Fig. S2B,C) while the concentration of traps near the nucleus declined (Fig. S2A). Thus, the barrier formed by the nuclear membrane is an important factor in trap generation.

**Polarity, not filament orientation, dominates transport times**

Since reversing the polarity of all filaments in a network with traps was sufficient to cause an overall decrease in MFPTs (Fig. 3E), we hypothesized that filament polarity could generally be utilized to regulate transport dynamics. We therefore fixed the orientations and positions of all filaments in a random network and reoriented all filament polarities such that they pointed outward (away from the nucleus). This global polarization resulted in an average 5-fold reduction in MFPT (Fig. 3F). In contrast, polarizing only 25% of the filaments in a random network such that they pointed inward (toward the nucleus) approximately doubled the average MFPT (Fig. 3G). Interestingly, average MFPT was not affected when we reoriented filament directions from a random network to one in which all filaments were radially aligned, keeping the polarities and center of masses of the filaments fixed (Fig. S3).

**Enhanced filament binding by multiple motors can have negative effects on transport efficiency**

Transport via multiple motors bound to a single cargo is common *in vivo* (35, 36) and has been the subject of numerous *in vitro* studies (37, 38). Previous studies (39, 40) modeled the effects of adding motors to cargo-motor complexes by increasing the on rate and decreasing the off rate relative to a single-motor scenario, increasing the time spent on the network during any binding event. To study how binding and unbinding rates affect MFPT statistics, we examined a wide range of values for random networks with 150 filaments of 3 μm in length, a set of network parameters that yields an intermediate MFPT (Fig. 2B). For high on and off rates, increases in either rate resulted in increases in MFPT (Fig. 4A), although MFPT also increased for small $k_{on}$ or $k_{off}$ below ~2/s (Fig. 4A), indicating that there is an optimal time spent on the network and processivity along filaments to minimize MFPT. MFPT variability over networks also increased substantially with increasing $k_{on}$, but decreased somewhat with increasing $k_{off}$ (Fig. 4B). For a collection of four motors bound to a cargo, each with a typical filament binding on rate of 5/s and an off rate of 0.005/step, a previous study, which simplistically assumed that the motors act completely independent of each other, predicted an effective collective motor-cargo on rate of 20/s and an off rate of 0.0001/step (39). Given these values, our simulations (Fig. 4) predict that the action of multiple motors could result in an average MFPT that is twice as long as the average MFPT in a scenario with just one motor, with a large increase in the variability of MFPTs across network realizations. More realistic models of multiple motor transport that address the effective interactions between processive motor proteins also predict the optimality of single motor transport based primarily on geometric and kinetic constraints (41, 42). Nonetheless, these models still predict an increase in the effective on rate and a decrease in the off rate for multiple motor cargos similar to the independent motor model studied, in which scenario our simulations show that the lowest MFPT for cargo transport is achieved by single processive motors given their optimal network residence time (Fig. 4).

# DISCUSSION

Given the importance of transport for many fundamental cellular processes, minimizing both the mean and variation in transport time from the nucleus to the cell membrane may result in increased cellular fitness. Our initial continuum model simulations (Fig. 1) suggest that the time of transport from the nucleus to the cell surface is minimized when transport systems with a constrained volume are localized adjacent to the nuclear surface in cells with nuclear radius greater than ~35% of the cell radius. *In vivo*, the ratio between the nuclear and cellular dimensions in most eukaryotic cells has been empirically observed to be 0.4-0.5 (25, 26). The physical mechanism underlying the minimum in transport time is due to the prediction from the diffusion equation for a circular geometry that cargo residence time due to transport via only diffusion through a cytoplasmic bulk is greatest at ~35% of the cell radius. Since the nuclear surface generally lies in or near this region of the cell, placing a transport system closer to the cell nucleus results in speeding up transport in the region where the cargo would otherwise spend more time, thus resulting in the greatest reduction in MFPT. Although many cellular, motor, and cytoskeletal network properties may affect transport efficiency, our simulations reveal that MFPT variability is a key quantity that is generally sensitive to specific details of the network architecture (Fig. 2C,3). High MFPTs relative to the average result from small subsets of filaments that form traps near the nucleus, the occurrence of which can be minimized in systems with a larger number of shorter filaments (Fig. 2C) and in systems with a buffer region between the nucleus and the transport network (Fig. S2). Our results suggest that in situations in which highly variable transport times can be detrimental to cellular function, eukaryotic cells may have evolved mechanisms to protect cargos against becoming trapped in the nucleus; these mechanisms could include establishing a buffer region around the nuclear region that lacks polymerized, random, cytoskeletal filament networks (Fig. S2). Relatively lower actin network densities near the nucleus have generally been reported, especially in the context of crawling cells (43, 44) as well as during nuclear positioning (45, 46). While in these cases different functional roles or causes for actin depletion have been postulated, our results suggest that an additional benefit of this arrangement is an avoidance of trapping in transport. Additionally, cellular filaments are likely to be dynamic; our transport results, which were obtained for fixed networks, can be understood as instantaneous characterizations of dynamic networks. Network dynamics and intracellular transport could be coupled *in vivo* (47), potentially allowing for feedback-regulated searches of the network parameter space resulting in beneficial outcomes such as MFPT reduction and responsiveness to changes in the external environment.

Since transport behavior is more strongly affected by filament polarity than by filament orientation (Fig. 3,S3), adding a few outwardly polarized filaments dramatically affected transport times (Fig. S4), suggesting a potential benefit of the microtubule nucleation at the nucleus that occurs in many eukaryotic cells. With regards to future experimental research with transport networks, these results can provide a framework for the measurement and exploration of *in vivo* networks properties that act to functionally reduce MFPT and MFPT variability.

In previous models (2, 5, 12, 15) that ignored the explicit spatial architecture of the network, the durations of each alternating phase of diffusive and ballistic motion during cargo transport were typically assumed to be exponentially distributed. However, in our simulations with explicit networks, the distribution of ballistic phases was complex and often strongly divergent from exponential (Fig. S5). Previous virtual network models could not yield information on variations in MFPT over network realizations, as the networks themselves were not explicitly modeled. Differences between our explicit physical modeling and previous virtual filament models can be substantial due to cargo rebinding and trapping effects, which can introduce errors in measuring MFPT of up to ~60% (Fig. S6).

Context is important when considering whether strong binding between the motor-cargo complex and the filament track is beneficial to the cell. For random networks, our simulations revealed that higher on rates and lower off rates than those typical for individual kinesin/myosin motors increase MFPT (Fig. 4A), suggesting that binding of cargos to multiple motors, which has been observed *in vivo* (35, 36), is not necessarily beneficial for transport. In these cases, periods of bulk diffusion enable cargos to disengage from regions of the transport network that are non-optimally oriented, such as traps near the nucleus (Fig. 3), and to explore neighboring regions of the cytoplasm. This result is consistent with recent multiple-motor models and experiments, which predict that processive motors are unlikely to collaborate, predominantly due to geometric and kinetic constraints (41, 42). However, along outwardly polarized filaments, stronger binding is likely to always decrease MFPT given that filaments point directly towards the target of transport.

Recently, experiments in vivo and in vitro have demonstrated the existence of anomalous features in intracellular transport (48-50), observing cargo motion as combinations of continuous time random walk (CTRW) and fractional Brownian-Langevin motion (fBM). Some of these experiments observed intracellular transport as primarily ergodic without aging (49, 51), with active network transport fundamentally modeled as an fBM process. Statistically, we find cargo motion in our simulations to be consistent with fBM, which is ergodic (50, 52). Firstly, the time-averaged mean squared displacement (MSD) for cargo transport as a function of lag time showed super-diffusive motion with an exponent α = 1.44 (where MSD~$t^\alpha$), which is consistent with the exponent measured with microspheres in cell extracts (49). Secondly, the ensemble-averaged MSD for cargo transport as a function of time shows convergence after ~100 s, implying that the system obeys ergodic behavior without aging (48). We do not observe a contribution from a CTRW-like process, which would be non-ergodic, wherein the ensemble-averaged MSD continues to decrease with time (48). Experiments that show aging and non-ergodicity are performed in vivo or with crowded cytosol extract, which in of itself can induce non-ergodic motion (48, 49), while our simulations are the result of a simple coupling of pure bulk diffusion and ballistic motion on fixed explicit filaments.

Overall, our results suggest a diverse range of mechanisms by which molecular transport over filaments can be tuned and regulated. We hypothesize that, in addition to reducing MFPT, reducing the variability in MFPT across network realizations promotes transport robustness over time. For example, having low variance in MFPT between network realizations among individual cells in a population, even at the expense of having a higher MFPT, is likely to promote fitness at the population level by virtue of having many fewer instances of cells with particularly poor network transport. Although we observed that MFPT was more sensitive to filament polarity (Fig. 3E-G) than to filament orientation (Fig. S4), using transport constraints to design the polarity of a network based on transport constraints may be challenging because a single network is used to transport cargos in multiple directions (9) and because cytoskeletal filaments also have other important structural roles (53). For biomimetic and other applications, however, we have demonstrated that several network properties, such as localization, filament length and number, and on and off rates, can be readily and rationally tuned.

# AUTHOR CONTRIBUTIONS

A.G., K.C.H., D.A., and N.C. designed the research; D.A., and N.C. performed the research; A.G., K.C.H., D.A., and N.C. analyzed the data; and A.G., K.C.H., and D.A. wrote the article.

# ACKNOWLEDGMENTS

We thank members of the Huang and Gopinathan labs for helpful discussions. This work was partially supported by National Science Foundation (NSF) grants EF-1038697 (to A.G. and K.C.H.) and DBI-0960480 (to A.G.), the Stanford Systems Biology Center funded by National Institutes of Health grant P50 GM107615 (to K.C.H.), and by a James S. McDonnell Foundation Award (to A.G.). This work was also supported in part by the National Science Foundation under Grant PHYS-1066293 and the hospitality of the Aspen Center for Physics.

# REFERENCES

1.  Arcizet, D., B. Meier, E. Sackmann, J. O. Radler, and D. Heinrich. 2008. Temporal Analysis of Active and Passive Transport in Living Cells. Phys Rev Lett 101.

2.  Bressloff, P., and J. M. Newby. 2013. Stochastic models of intracellular transport. Reviews of Modern Physics 85.

3.  Ross, J. L., M. Y. Ali, and D. M. Warshaw. 2008. Cargo transport: molecular motors navigate a complex cytoskeleton. Curr Opin Cell Biol 20:41-47.

4.  Wells, A. L., A. W. Lin, L. Q. Chen, D. Safer, S. M. Cain, T. Hasson, B. I. Carragher, R. A. Milligan, and H. L. Sweeney. 1999. Myosin VI is an actin-based motor that moves backwards. Nature 401:505-508.

5.  Loverdo, C., O. Benichou, M. Moreau, and R. Voituriez. 2008. Enhanced reaction kinetics in biological cells. Nat Phys 4:134-137.

6.  Langford, G. M. 1995. Actin-Dependent and Microtubule-Dependent Organelle Motors - Interrelationships between the 2 Motility Systems. Curr Opin Cell Biol 7:82-88.

7.  Wang, Z., S. Khan, and M. P. Sheetz. 1995. Single cytoplasmic dynein molecule movements: characterization and comparison with kinesin. Biophysical journal 69:2011-2023.

8.  Mallik, R., and S. P. Gross. 2004. Molecular Motors: Strategies to Get Along. Current Biology 14:R971-R982.

9.  Howard, J. 2001. Mechanics of motor proteins and the cytoskeleton.

10. Snider, J., F. Lin, N. Zahedi, V. Rodionov, C. C. Yu, and S. P. Gross. 2004. Intracellular actin-based transport: How far you go depends on how often you switch. P Natl Acad Sci USA 101:13204-13209.

11. Rodionov, V., J. Yi, A. Kashina, A. Oladipo, and S. P. Gross. 2003. Switching between microtubule- and actin-based transport systems in melanophores is controlled by cAMP levels. Curr Biol 13:1837-1847.

12. Bressloff, P., and J. Newby. 2009. Directed intermittent search for hidden targets. New Journal of Physics 11.

13. Kuznetsov, A. V., A. A. Avramenko, and D. G. Blinov. 2008. Numerical modeling of molecular-motor-assisted transport of adenoviral vectors in a spherical cell. Computer methods in biomechanics and biomedical engineering 11:215-222.

14. Smith, D. A., and R. M. Simmons. 2001. Models of motor-assisted transport of intracellular particles. Biophysical journal 80:45-68.

15. Condamin, S., O. Benichou, V. Tejedor, R. Voituriez, and J. Klafter. 2007. First-passage times in complex scale-invariant media. Nature 450:77-80.

16. Kahana, A., G. Kenan, M. Feingold, M. Elbaum, and R. Granek. 2008. Active transport on disordered microtubule networks: the generalized random velocity model. Physical review. E, Statistical, nonlinear, and soft matter physics 78:051912.

17. Tseng, Y., T. P. Kole, and D. Wirtz. 2002. Micromechanical mapping of live cells by multiple-particle-tracking microrheology. Biophysical journal 83:3162-3176.

18. McVicker, D. P., L. R. Chrin, and C. L. Berger. 2011. The nucleotide-binding state of microtubules modulates kinesin processivity and the ability of Tau to inhibit kinesin-mediated transport. The Journal of biological chemistry 286:42873-42880.

19. Alberts, B. 2008. Molecular biology of the cell. Garland Science, New York.

20. Ham, R. G. 1965. Clonal Growth of Mammalian Cells in a Chemically Defined Synthetic Medium. P Natl Acad Sci USA 53:288-&.

21. Xu, J., S. J. King, M. Lapierre-Landry, and B. Nemec. 2013. Interplay between velocity and travel distance of kinesin-based transport in the presence of tau. Biophysical journal 105:L23-25.

22. Mehta, A. D., R. S. Rock, M. Rief, J. A. Spudich, M. S. Mooseker, and R. E. Cheney. 1999. Myosin-V is a processive actin-based motor. Nature 400:590-593.

23. Reck-Peterson, S. L., A. Yildiz, A. P. Carter, A. Gennerich, N. Zhang, and R. D. Vale. 2006. Single-molecule analysis of dynein processivity and stepping behavior. Cell 126:335-348.

24. Reilein, A., S. Yamada, and W. J. Nelson. 2005. Self-organization of an acentrosomal microtubule network at the basal cortex of polarized epithelial cells. The Journal of cell biology 171:845-855.

25. Huber, M. D., and L. Gerace. 2007. The size-wise nucleus: nuclear volume control in eukaryotes. The Journal of cell biology 179:583-584.

26. Neumann, F. R., and P. Nurse. 2007. Nuclear size control in fission yeast. The Journal of cell biology 179:593-600.

27. Schulze, E., and M. Kirschner. 1986. Microtubule Dynamics in Interphase Cells. Journal of Cell Biology 102:1020-1031.

28. Soltys, B. J., and G. G. Borisy. 1985. Polymerization of tubulin in vivo: direct evidence for assembly onto microtubule ends and from centrosomes. The Journal of cell biology 100:1682-1689.

29. Boyles, J., J. E. Fox, D. R. Phillips, and P. E. Stenberg. 1985. Organization of the cytoskeleton in resting, discoid platelets: preservation of actin filaments by a modified fixation that prevents osmium damage. The Journal of cell biology 101:1463-1472.

30. Hartwig, J. H., and M. DeSisto. 1991. The cytoskeleton of the resting human blood platelet: structure of the membrane skeleton and its attachment to actin filaments. The Journal of cell biology 112:407-425.

31. Xu, K., H. P. Babcock, and X. Zhuang. 2012. Dual-objective STORM reveals three-dimensional filament organization in the actin cytoskeleton. Nature methods 9:185-188.

32. Vale, R. D., and R. A. Milligan. 2000. The way things move: looking under the hood of molecular motor proteins. Science 288:88-95.

33. Dragovic, R. A., C. Gardiner, A. S. Brooks, D. S. Tannetta, D. J. Ferguson, P. Hole, B. Carr, C. W. Redman, A. L. Harris, P. J. Dobson, P. Harrison, and I. L. Sargent. 2011. Sizing and phenotyping of cellular vesicles using Nanoparticle Tracking Analysis. Nanomedicine : nanotechnology, biology, and medicine 7:780-788.

34. Medalia, O., D. Typke, R. Hegerl, M. Angenitzki, J. Sperling, and R. Sperling. 2002. Cryoelectron microscopy and cryoelectron tomography of the nuclear pre-mRNA processing machine. Journal of structural biology 138:74-84.

35. Levi, V., A. S. Serpinskaya, E. Gratton, and V. Gelfand. 2006. Organelle transport along microtubules in Xenopus melanophores: evidence for cooperation between multiple motors. Biophysical journal 90:318-327.

36. Welte, M. A., S. P. Gross, M. Postner, S. M. Block, and E. F. Wieschaus. 1998. Developmental regulation of vesicle transport in Drosophila embryos: forces and kinetics. Cell 92:547-557.

37. Vershinin, M., B. C. Carter, D. S. Razafsky, S. J. King, and S. P. Gross. 2007. Multiple-motor based transport and its regulation by Tau. Proc Natl Acad Sci U S A 104:87-92.

38. Ando, D., M. K. Mattson, J. Xu, and A. Gopinathan. 2014. Cooperative protofilament switching emerges from inter-motor interference in multiple-motor transport. Scientific reports 4:7255.

39. Klumpp, S., and R. Lipowsky. 2005. Cooperative cargo transport by several molecular motors. Proc Natl Acad Sci U S A 102:17284-17289.

40. Beeg, J., S. Klumpp, R. Dimova, R. S. Gracia, E. Unger, and R. Lipowsky. 2008. Transport of beads by several kinesin motors. Biophysical journal 94:532-541.

41. Driver, J. W., D. K. Jamison, K. Uppulury, A. R. Rogers, A. B. Kolomeisky, and M. R. Diehl. 2011. Productive Cooperation among Processive Motors Depends Inversely on Their Mechanochemical Efficiency. Biophysical journal 101:386-395.

42. Kolomeisky, A. B. 2013. Motor proteins and molecular motors: how to operate machines at the nanoscale. J Phys-Condens Mat 25.

43. Lewalle, A., M. Fritzsche, K. Wilson, R. Thorogate, T. Duke, and G. Charras. 2014. A phenomenological density-scaling approach to lamellipodial actin dynamics(dagger). Interface focus 4:20140006.

44. Small, J. V., M. Herzog, and K. Anderson. 1995. Actin filament organization in the fish keratocyte lamellipodium. The Journal of cell biology 129:1275-1286.

45. Almonacid, M., W. W. Ahmed, M. Bussonnier, P. Mailly, T. Betz, R. Voituriez, N. S. Gov, and M. H. Verlhac. 2015. Active diffusion positions the nucleus in mouse oocytes. Nature cell biology 17:470-479.

46. Dupin, I., Y. Sakamoto, and S. Etienne-Manneville. 2011. Cytoplasmic intermediate filaments mediate actin-driven positioning of the nucleus. Journal of cell science 124:865-872.

47. Hehnly, H., and M. Stamnes. 2007. Regulating cytoskeleton-based vesicle motility. FEBS letters 581:2112-2118.

48. Tabei, S. M. A., S. Burov, H. Y. Kim, A. Kuznetsov, T. Huynh, J. Jureller, L. H. Philipson, A. R. Dinner, and N. F. Scherer. 2013. Intracellular transport of insulin granules is a subordinated random walk. P Natl Acad Sci USA 110:4911-4916.

49. Regner, B. M., D. Vucinic, C. Domnisoru, T. M. Bartol, M. W. Hetzer, D. M. Tartakovsky, and T. J. Sejnowski. 2013. Anomalous Diffusion of Single Particles in Cytoplasm. Biophysical journal 104:1652-1660.

50. Jeon, J. H., V. Tejedor, S. Burov, E. Barkai, C. Selhuber-Unkel, K. Berg-Sorensen, L. Oddershede, and R. Metzler. 2011. In vivo anomalous diffusion and weak ergodicity breaking of lipid granules. Phys Rev Lett 106:048103.

51. Weiss, M. 2013. Single-particle tracking data reveal anticorrelated fractional Brownian motion in crowded fluids. Physical review. E, Statistical, nonlinear, and soft matter physics 88:010101.

52. Duan, H. G., and X. T. Liang. 2012. Ergodic properties of fractional Langevin motion with spatial correlated noise. Eur Phys J B 85.

53. Appaix, F., A. V. Kuznetsov, Y. Usson, L. Kay, T. Andrienko, J. Olivares, T. Kaambre, P. Sikk, R. Margreiter, and V. Saks. 2003. Possible role of cytoskeleton in intracellular arrangement and regulation of mitochondria. Experimental physiology 88:175-190.

54. Luby-Phelps, K. 2000. Cytoarchitecture and physical properties of cytoplasm: volume, viscosity, diffusion, intracellular surface area. International review of cytology 192:189-221.

**Figure legends**

**Figure 1: Localization of a cytoskeletal network near the nuclear surface minimizes transport time to the membrane in a continuum diffusion model.** A) In these simulations, cargo-motor complexes move through the cytoplasm, which is modeled by bulk (blue) and network (green) regions. Our simulations are typically initialized with cargo on the surface of an impenetrable nucleus (pink), and transport ends when the cargo reaches the cell membrane (red circle). The cytoskeletal network is represented by an annulus of width $w$ located at an inner radius $R_a$ within which the diffusion constant is increased to $D_a = 1.1$ µm²/s from $D = 0.011$ µm²/s in the bulk (54). The annulus is placed outside of a nucleus with radius $R_n$. B) Average cargo residence time peaked ~3.5 µm from the cell center for simulations of 10,000 cargos that terminate transport at the cell membrane in a cell with no nucleus ($R_n = 0$) and only bulk diffusion with constant $D$. Grey line, analytical calculation of residence times (see Supporting Material). C) Mean first passage time (MFPT) after addition of an annulus with diffusion constant $D_a = 1.1$ µm²/s, width $w = 3$ µm. The annulus position was shifted from the nuclear surface to the cell membrane, and each result is the mean over 1,000 cargos. Black lines indicate analytically determined residence times. For $R_n \lesssim 2$ µm, MFPT was minimized at an intermediate $R_a$; for larger $R_n$, MFPT was minimized with the annulus positioned adjacent to the nucleus.

**Figure 2: MFPT is conserved with respect to total filament mass, but MFPT variability increases with fewer filaments.** A) Example of an explicit-filament representation of a cytoskeletal network for simulations in which cargo-motor complexes diffuse through the cytoplasmic cell bulk or move ballistically along transport filaments. Complexes are assumed to

originate at the nuclear surface and transport terminates at the cell membrane. Red and green circles represent the (+) and (-) ends of randomly oriented filaments, respectively. The white circle indicates a 0.2-µm buffer between the nucleus and the filaments. B) MFPTs in explicit networks with a range of filament lengths and numbers; $v = 1$ µm/s, $D = 0.051$ µm$^2$/s, $k_{on} = 5$/s, $k_{off} = 1$/s. MFPT is approximately constant along the black lines, which represent conserved network mass. C) The standard deviation in MFPT over network realizations, normalized by MFPT, typically increases for smaller numbers of longer filaments.

**Figure 3: High MFPTs result from traps near the nucleus.** A) Histograms of average MFPTs for 3,000 cargos for each of 500,000 network realizations consisting of random networks of 150 filaments with length 3.5 µm (pink) and 500 filaments with length 4.5 µm (blue). B) Cargo residence times reveal traps near the nuclear surface for a network with a relatively large MFPT (220 s) relative to the mean over all networks (82 s) from the pink distribution in part (A). C) Cargo residence times of the network in (B) with the polarity of all filaments reversed; this reversal yielded a below-average MFPT of 62 s. D) Total network trap area is strongly connected with MFPT across networks. E) Reversing the polarity of all filaments in a network decreases the MFPT of slow networks. F) Polarity reversal for 100% of the filaments that previously faced inward dramatically decreases MFPT. G) The MFPT of a network typically increases substantially after polarity reversal for 25% of the filaments that previously faced outward. All simulations in (D-G) were carried out with 3,000 cargos on 1,000 networks of 150 3.5-µm filaments. Red lines in (E-G) indicate the line $y=x$.

**Figure 4: MFPT and variability over networks are minimized for intermediate on rate.** A) MFPT for random networks of 150 filaments with length 3 μm, measured for 3,000 motors on 60,000 network realizations for each combination of $k_{on}$ and $k_{off}$, has a minimum at a motor on rate of $k_{on} \sim 5$/s. A motor step length of 8 nm, corresponding to the distance between subunits on a microtubule filament, was assumed; larger step length magnitudes merely result in a rescaling of the motor off rates in per step terms. B) Standard deviation of MFPTs across network realizations normalized by average MFPT decreases with decreasing on rate and increasing off rate.

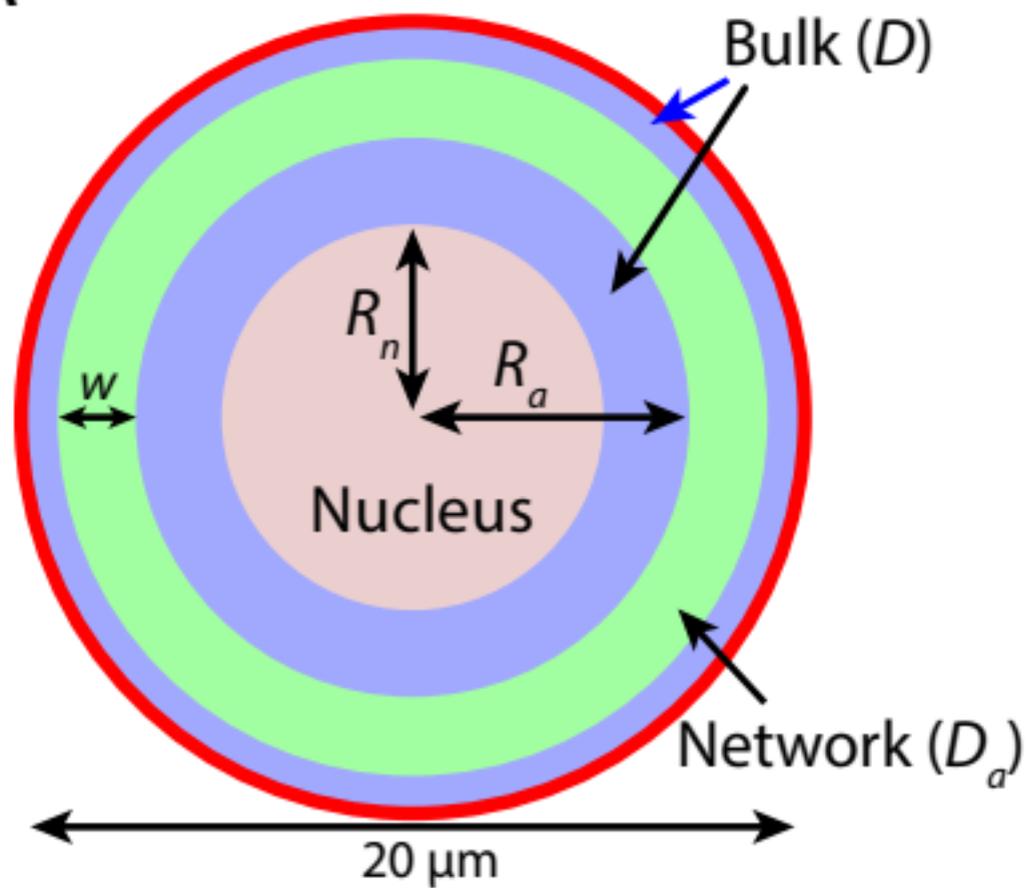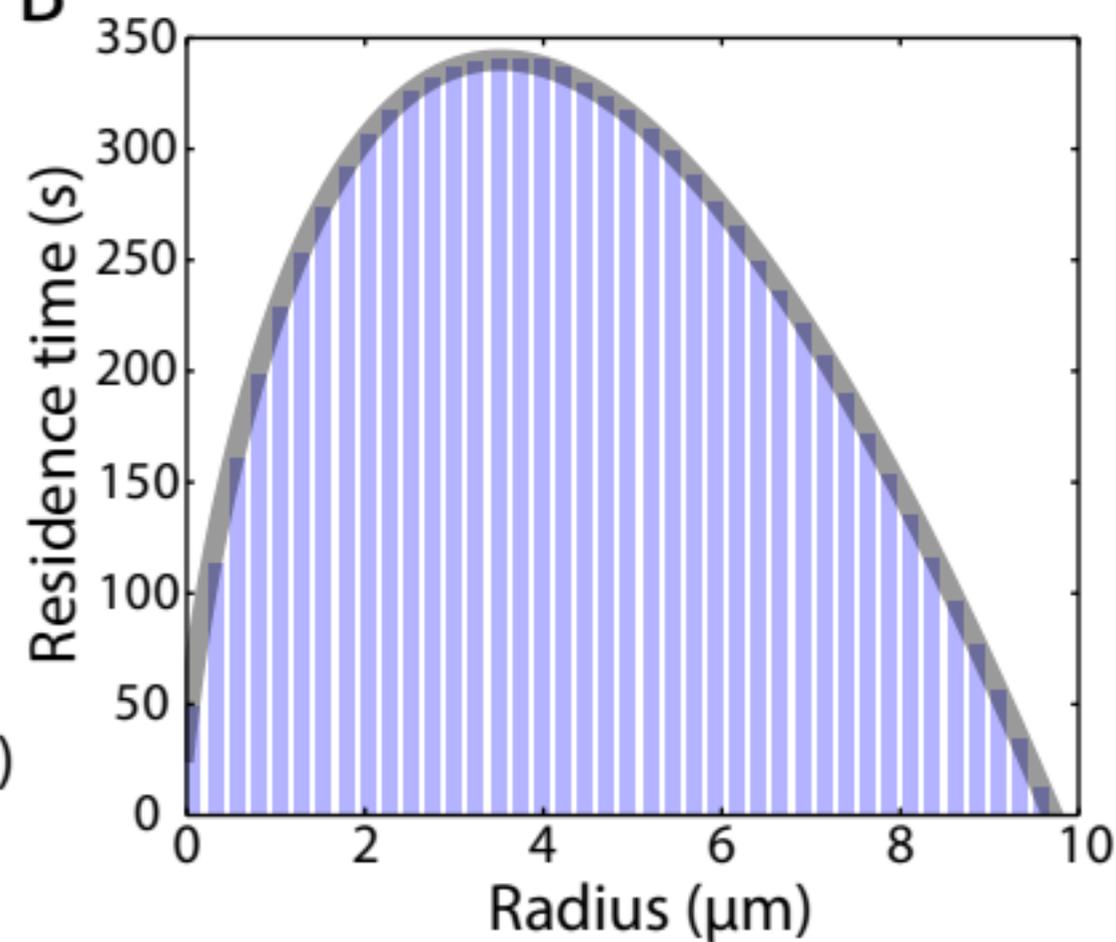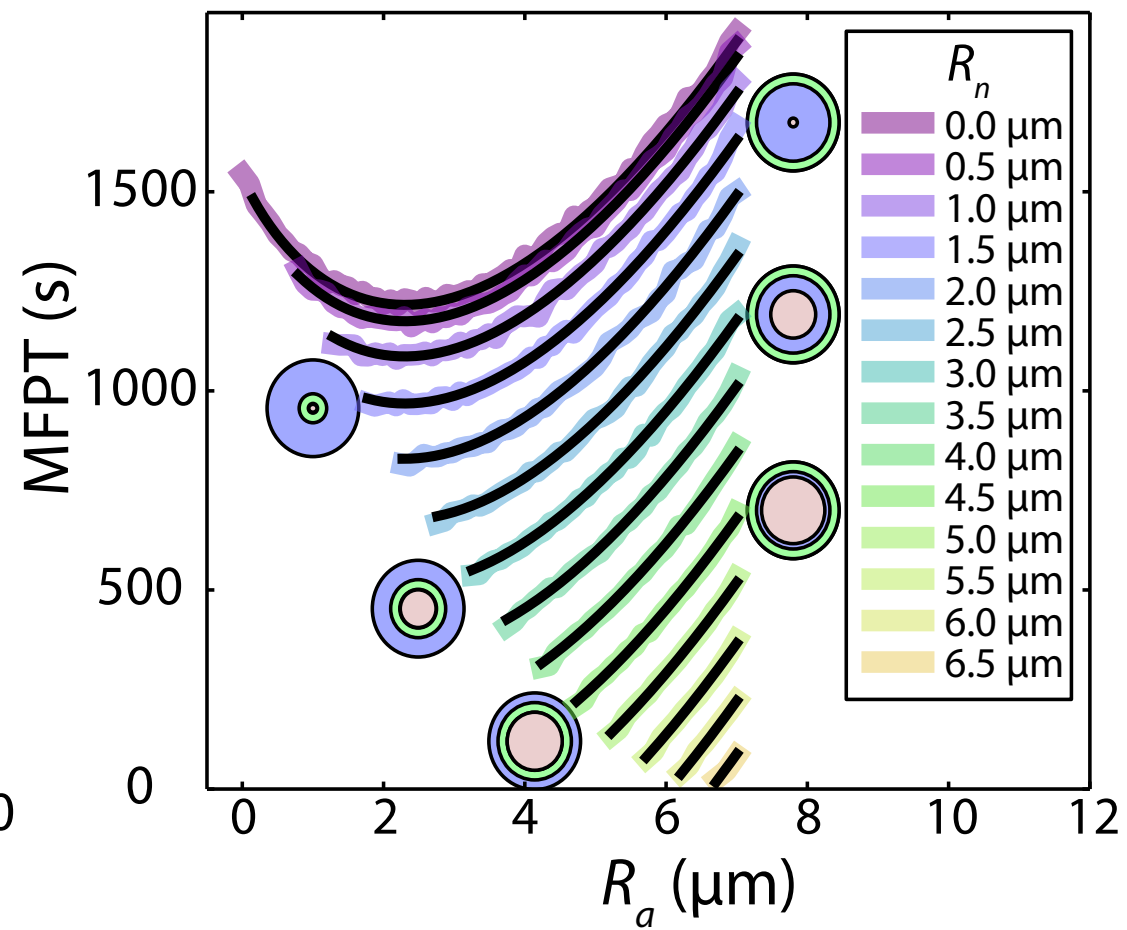

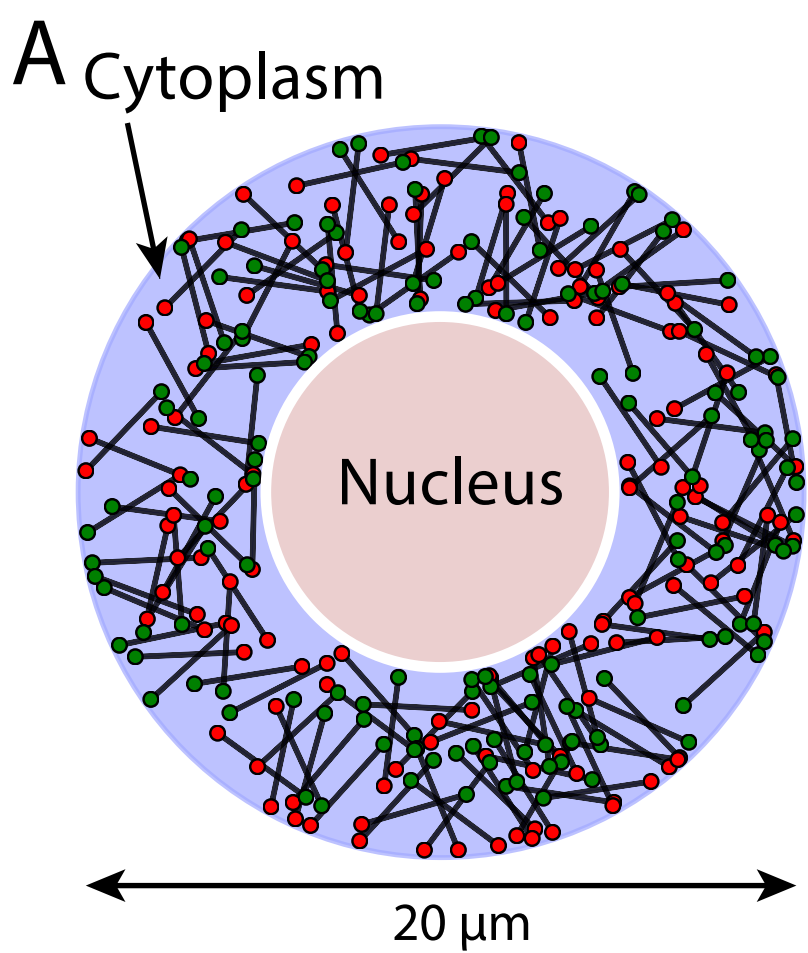 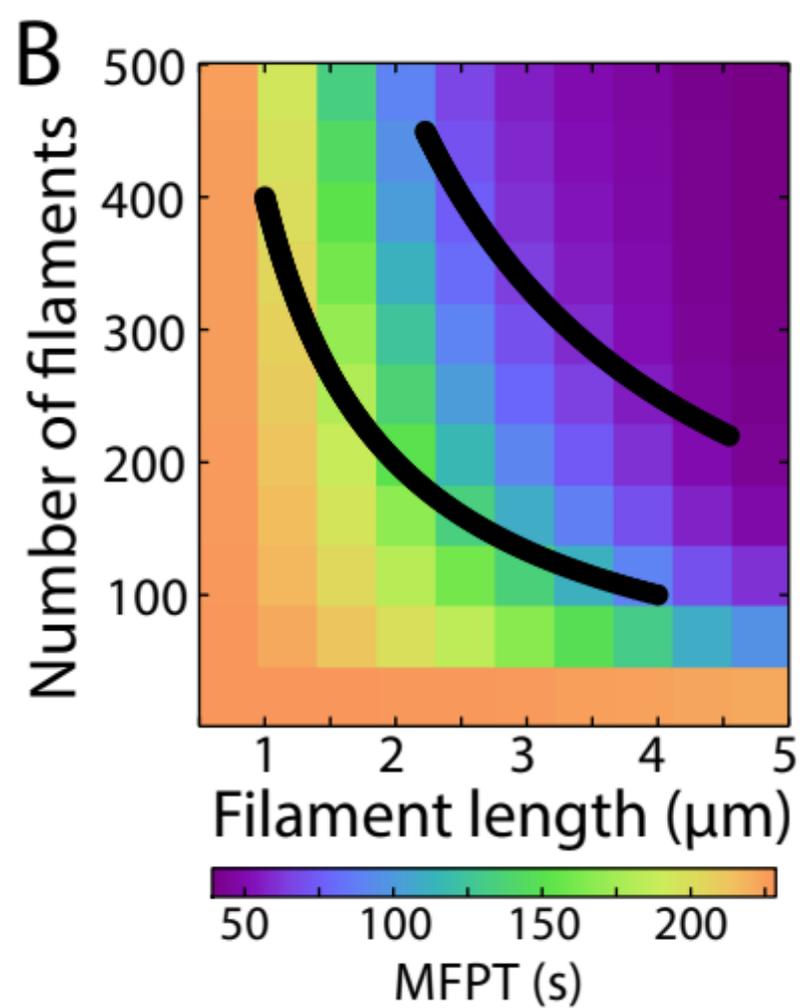 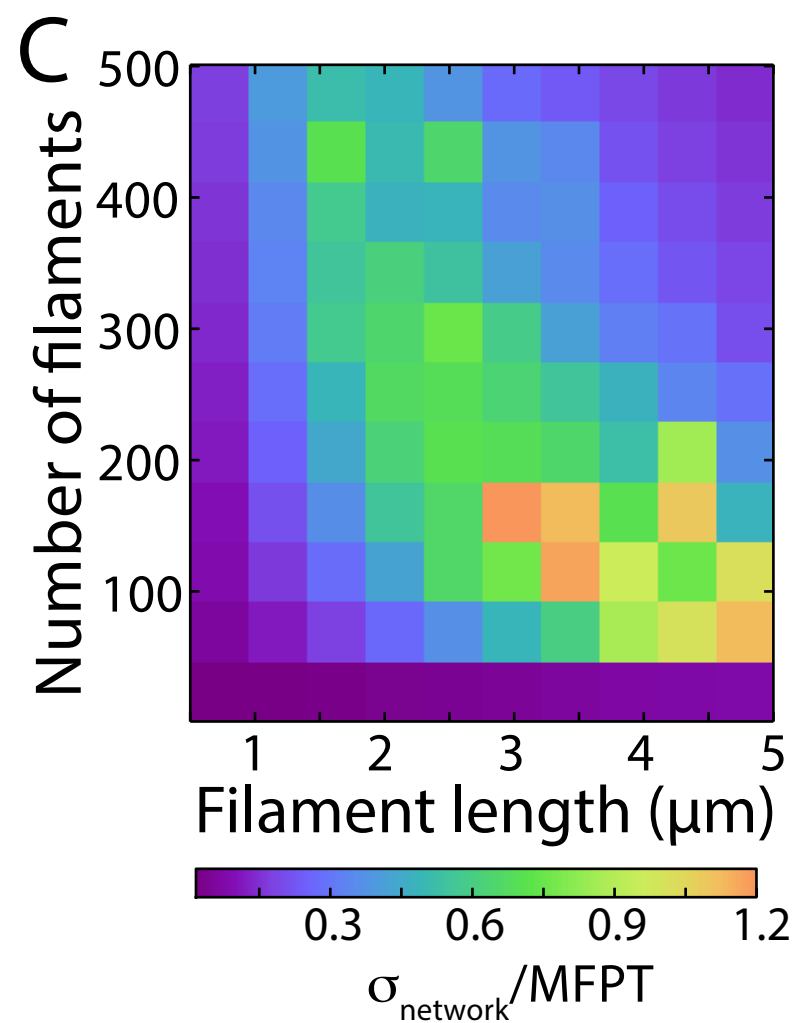

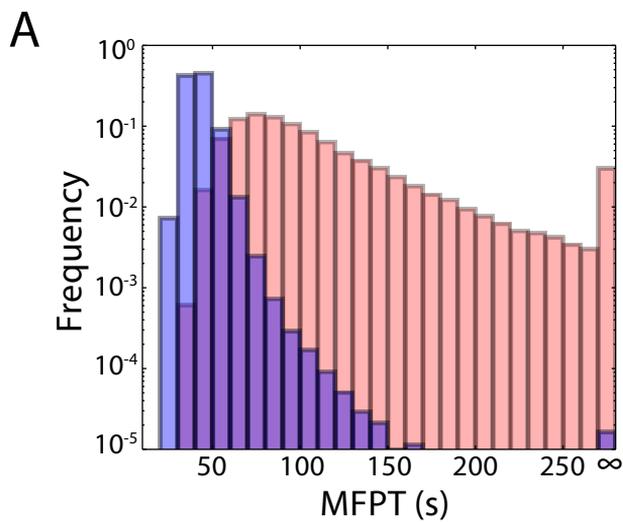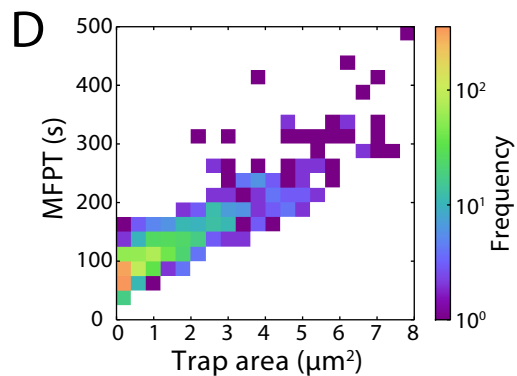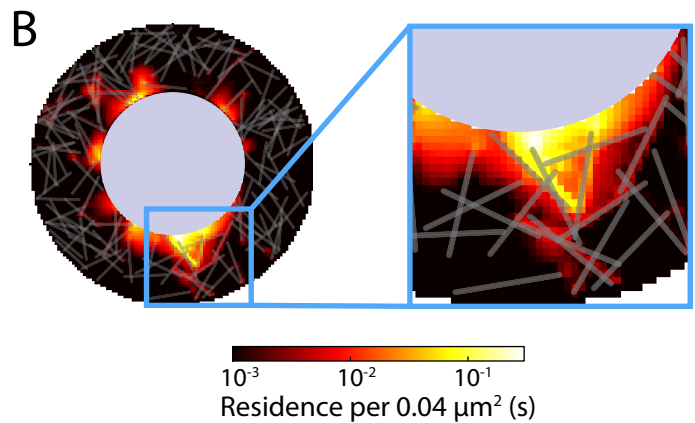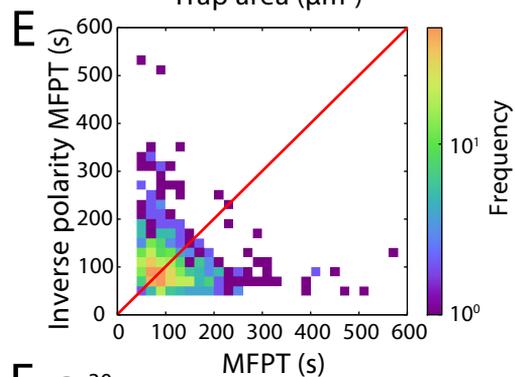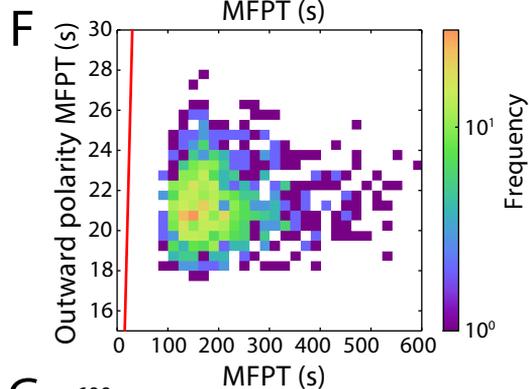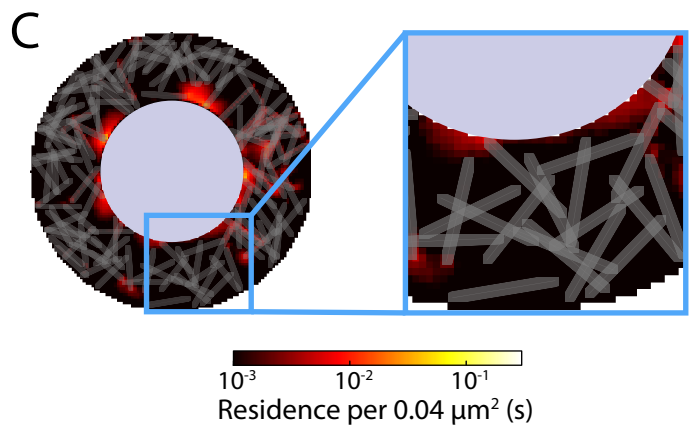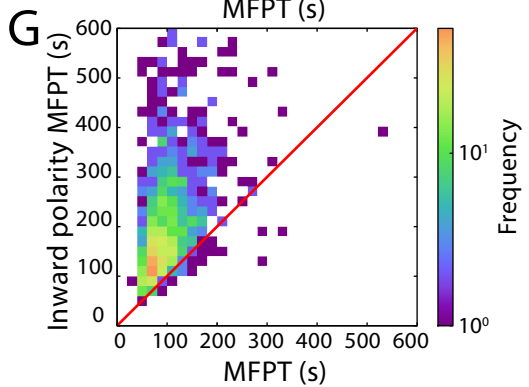

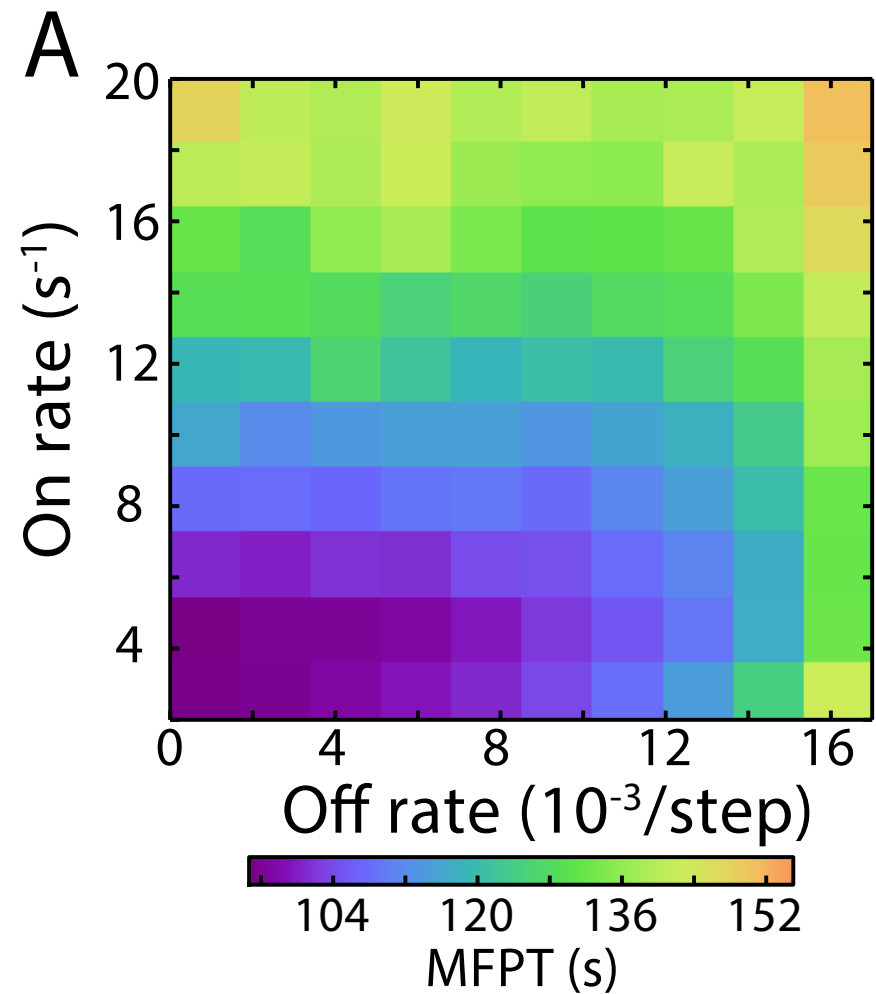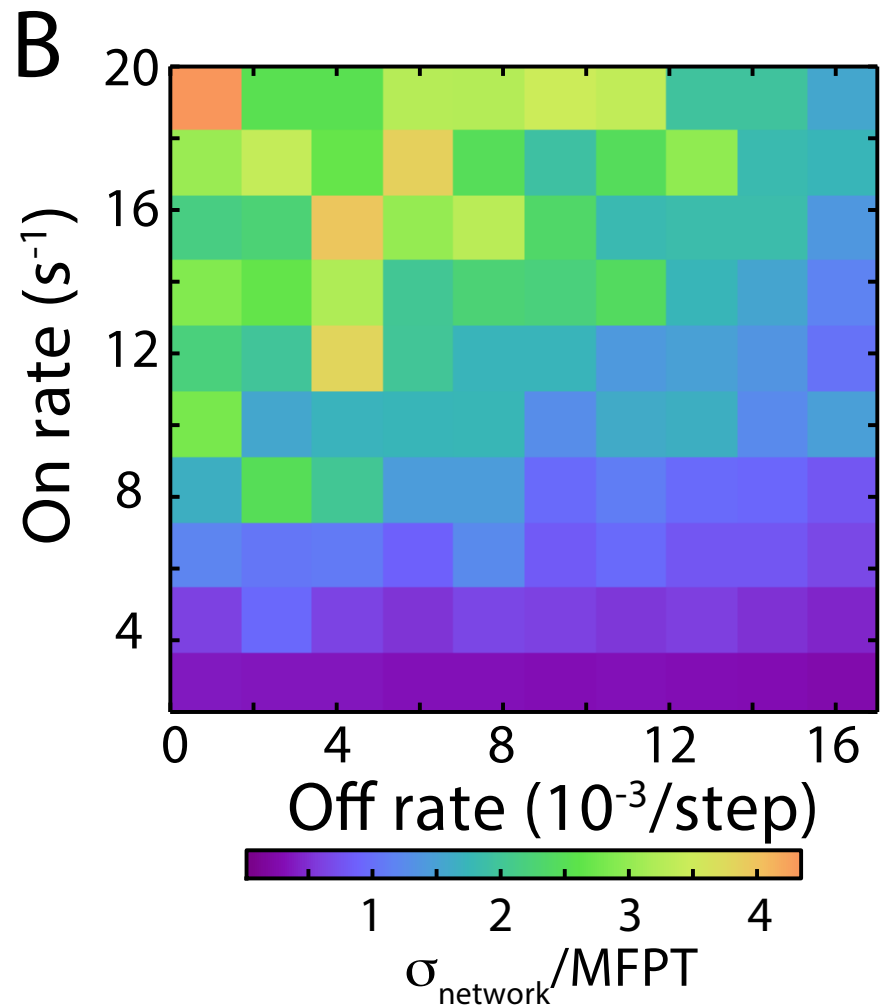

**Supplementary Information for "Cytoskeletal network morphology regulates intracellular transport dynamics"**

David Ando[1], Nickolay Korabel[1,2], Kerwyn Casey Huang[3,4,*], Ajay Gopinathan[1,*]

[1]Department of Physics, University of California at Merced, Merced, CA, USA

[2]School of Mathematics, University of Manchester, Manchester, UK

[3]Department of Bioengineering, Stanford University, Stanford, CA, USA

[4]Department of Microbiology and Immunology, Stanford University School of Medicine, Stanford, CA, USA

*Correspondence to: kchuang@stanford.edu, agopinathan@ucmerced.edu

## *Analytical solution of 2D diffusion equation*

The equation describing the concentration of cargos $c(x,y,t)$ diffusing in two dimensions is

$$\frac{\partial c}{\partial t} = D\left(\frac{\partial^2 c}{\partial x^2} + \frac{\partial^2 c}{\partial y^2}\right). \qquad (1)$$

For the transport problems we consider, we assume a constant source at either the center of the cell or at the nuclear boundary, and zero cargo concentration along the cell boundary.

At steady state, Eq. (1) transformed into polar coordinates becomes

$$D\left(\frac{\partial^2 c}{\partial x^2} + \frac{\partial^2 c}{\partial y^2}\right) = \left(\frac{\partial^2 c}{\partial r^2} + \frac{1}{r}\frac{\partial c}{\partial r} + \frac{1}{r^2}\frac{\partial^2 c}{\partial \theta^2}\right) = 0. \qquad (2)$$

Since the system is symmetric with respect to $\theta$, the solution is

$$c(r,\theta) = C_1 + C_2 \log r. \qquad (3)$$

The residence time of cargos within the interval $[r, r+dr]$ is $T(r,\theta)dr = 2\pi r c(r,\theta) dr$.

For cargos starting at $r=0$, the time required to diffusion a distance $R$, the radius of the cell, is $R^2/4D$, which should be equivalent to the integral of $T(r,\theta)$ from 0 to $R$:

$$\int_0^R 2\pi r(C_1 + C_2 \log r) dr = \frac{\pi R^2}{2}(2C_1 - C_2 + 2C_2 \log R) = \frac{R^2}{4D}. \qquad (4)$$

The boundary condition $c(R) = 2\pi R(C_1 + C_2 \log R) = 0$ specifies solutions for $C_1$ and $C_2$:

$$C_1 = \frac{\log R}{2\pi D}, \qquad (5)$$

$$C_2 = -\frac{1}{2\pi D}. \qquad (6)$$

## *Analytical solution of 2D diffusion equation including annulus with increased rate of diffusion*

For the simulations in Fig. 1C that include an annulus representing faster diffusion on the network, we consider cargos transported from the nuclear boundary at $R_n$ to the cell membrane at $R$ through the bulk with diffusion constant $D_1$ or through an annulus of width $w = 3$ μm located at a radius $R_a$ with diffusion constant $D_2$. The MFPT is simply the sum of integrals of $T(r,\theta)$ as solved above for different conditions:

$$\text{MFPT} = \int_{R_n}^{R_a} 2\pi r \left(\frac{\log R}{2\pi D_1} - \frac{\log r}{2\pi D_1}\right) dr + \int_{R_a}^{R_a+w} 2\pi r \left(\frac{\log R}{2\pi D_2} - \frac{\log r}{2\pi D_2}\right) dr + \int_{R_a+w}^{R_m} 2\pi r \left(\frac{\log R}{2\pi D_1} - \frac{\log r}{2\pi D_1}\right) dr. \quad (7)$$

After integration, this yields:

$$\text{MFPT} = \frac{(R_a^2 - 4)\log R - 2\left(1 - \log 4 + \frac{1}{4}R_a^2 (2\log R_a - 1)\right)}{2D_1} + \frac{2R_a^2 \log R_a - 2(3+R_a)^2 \log(3+R_a) + 3(3+2R_a)(1+2\log R)}{4D_2} +$$

$$\frac{(R^2 - (3+R_a)^2)\log R + 2\left(\frac{1}{4}(3+R_a)^2 (2\log(3+R_a) - 1) - \frac{1}{4}R^2 (2\log R - 1)\right)}{2D_1}. \quad (8)$$

For the simulations in Fig. S1 that include an annulus representing faster diffusion on the network, we consider cargos transported from the nuclear boundary at $R_n$ to the cell membrane at $R$ through the bulk with diffusion constant $D_1$ or through an annulus of fixed area $A = 15.6$ μm² located at a radius $R_a$ with diffusion constant $D_2$. The MFPT is again simply the sum of integrals of $T(r,\theta)$ as solved above for different conditions:

$$\text{MFPT} = \int_{R_n}^{R_a} 2\pi r \left(\frac{\log R}{2\pi D_1} - \frac{\log r}{2\pi D_1}\right) dr + \int_{R_a}^{\sqrt{R_a^2 + A}} 2\pi r \left(\frac{\log R}{2\pi D_2} - \frac{\log r}{2\pi D_2}\right) dr + \int_{\sqrt{R_a^2 + A}}^{R_m} 2\pi r \left(\frac{\log R}{2\pi D_1} - \frac{\log r}{2\pi D_1}\right) dr.$$

(9)

After integration, this yields:

$$\text{MFPT} = \frac{2\left(\frac{1}{4}R_a^2(2\log R_a - 1) - \frac{1}{4}(A+R_a^2)\left(\log(A+R_a^2)-1\right)\right) + A\log R}{4D_2} +$$

$$\frac{(R^2 - A - R_a^2)\log R - 2\left(-\frac{1}{4}(A+R_a^2)(\log(A+R^2)-1) + \frac{1}{4}R^2\,(2\log R - 1)\right)}{2D_1} + \frac{(R_a^2 - R_n^2)\log R + \frac{1}{2}(R_a^2 - R_n^2 - 2R_a^2 \log R_a + 2\,R_n^2 \log R_n)}{2D_1}.$$

(10)

The black curves in Figs. 1C and S1 are the analytical solutions for different values of $R_n$ and $R_a$ from Eqs. 8 and 10, respectively.

**Supplementary Figure Legends**

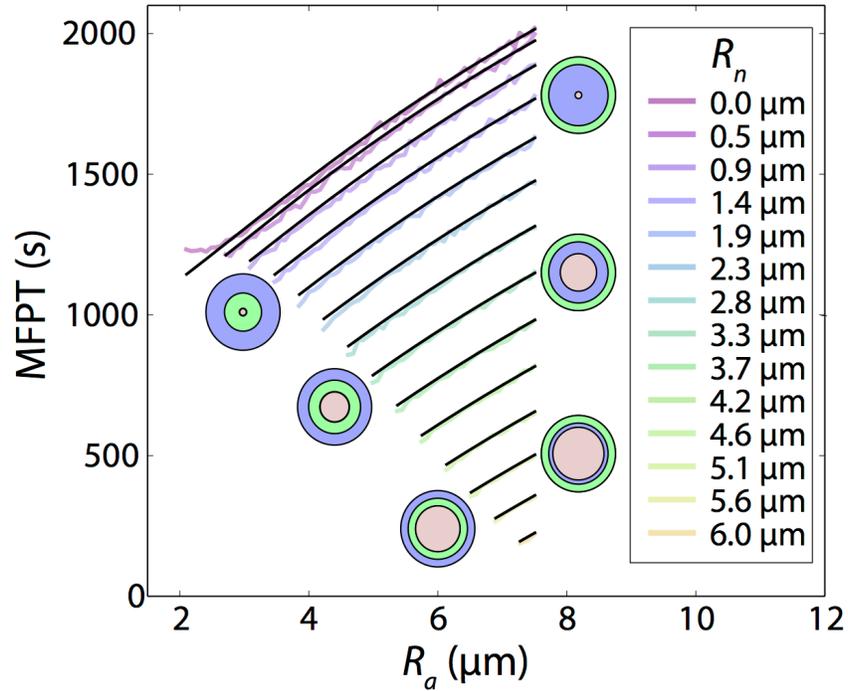

**Figure S1: MFPT increases monotonically as the position of an annulus of increased rate of diffusion with fixed area moves toward the cell membrane.** Within the annulus, the diffusion constant is $D_a = 1.1\ \mu m^2/s$, as compared to the lower diffusion with $D = 0.011\ \mu m^2/s$ in the rest of the cell. The area of the annulus is $A = 15.6\ \mu m^2$, and the different curves represent different nuclear radii $R_n$. Results were averaged over 1,000 cargos for each value of $R_n$ and $R_a$. The minimum MFPT is always achieved by placing the annulus as close to the nucleus as possible. Black curves are analytic predictions based on Eq. 10 in the SI.

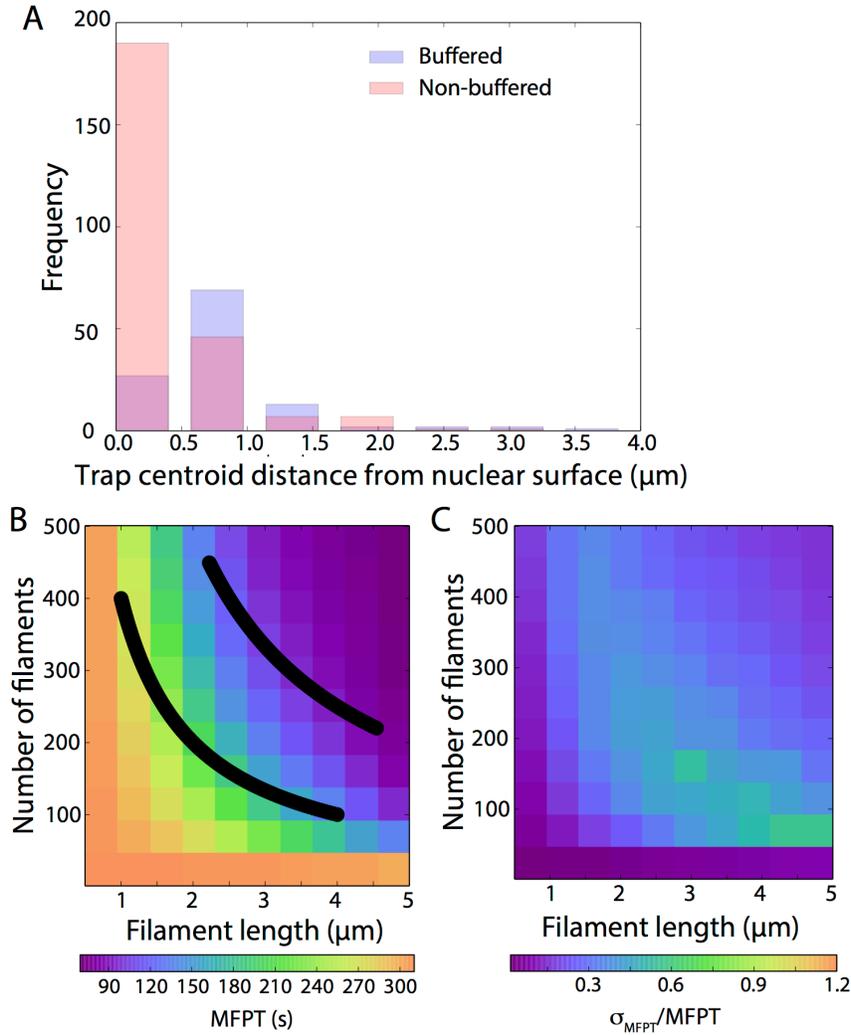

**Figure S2: A buffer region around the nucleus decreases both the incidence of traps and MFPT variability.** Explicit filament simulations in which a larger buffer (1 μm) was introduced between the filaments and nucleus by reducing the nuclear radius. A) The radial frequency of trap centroids for 10,000 non-buffered (pink) and buffered (blue) network realizations containing 150 3.5-μm filaments. Buffered simulations have reduced trap occurrence and reduced trap formation near the nuclear surface. B) MFPT, ignoring the passive transport time within the buffer, was relatively unaffected by the increased size of the buffer region (compare with Fig. 2B). C) The standard deviation of MFPTs across network realizations was substantially reduced by the increased size of the buffer region (compare with Fig. 2C).

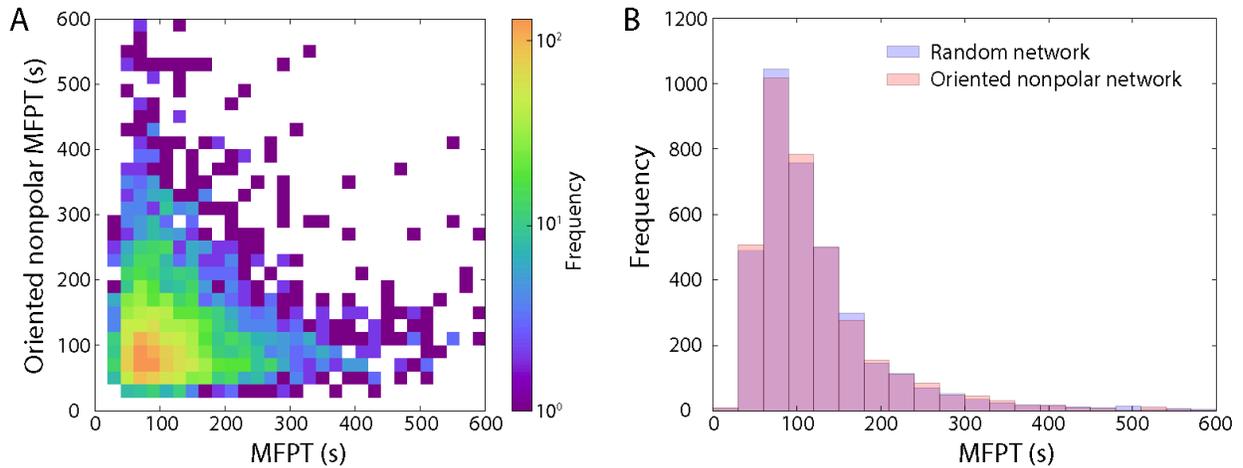

**Figure S3: Radially orienting filaments does not strongly affect MFPT.** MFPTs from simulations with random networks of 400 filaments with length 2 μm, compared to similar networks in which all filaments have been rotated counterclockwise until they are radially oriented. The polarity of each filament toward the nucleus or membrane is preserved. Simulations were performed for 3,500 network realizations. A) The distribution of MFPTs was similar in both cases, with a mean MFPT over networks of 143 s and 136 s for random and radially oriented networks, respectively. Given the non-diagonal nature of this scatter plot, the correlation between a given random network's MFPT and its rotated oriented nonpolar network MFPT is low. B) MFPT distributions for random and oriented network realizations are nearly identical.

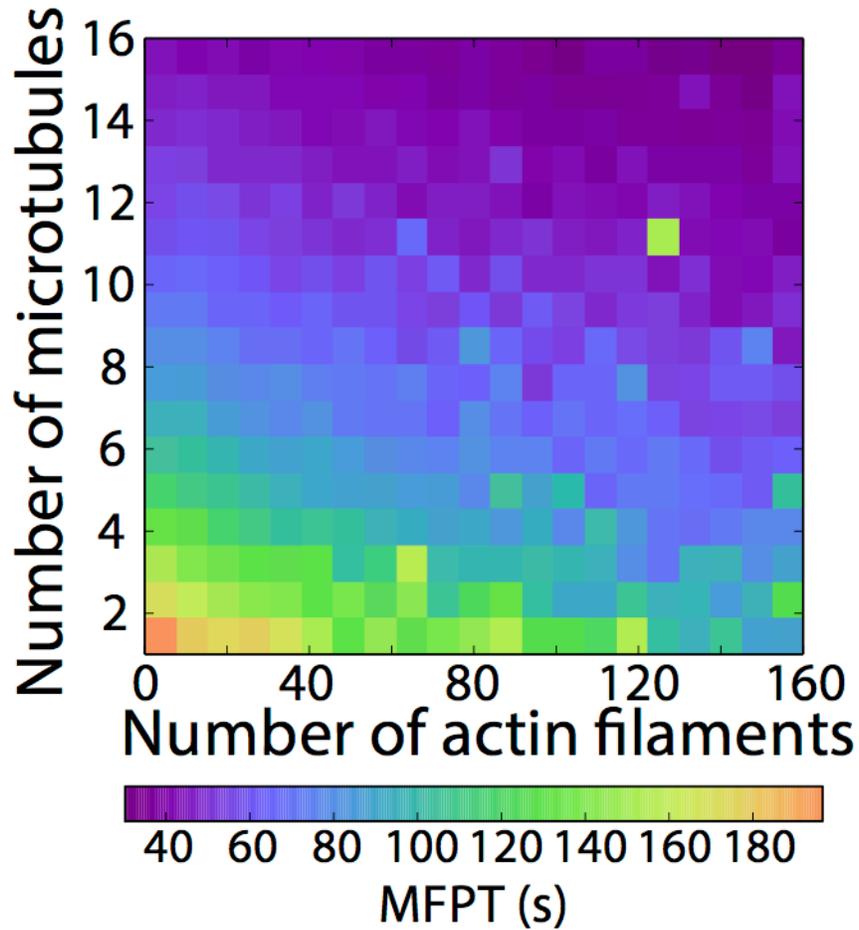

**Figure S4: MFPT decreases dramatically when a small number of outwardly polarized radial filaments are superimposed on a random filament network.** Actin filaments (3 µm in length) were randomly distributed, oriented, and polarized as in Figs. 2,3. Microtubules (5 µm in length) were assumed to be nucleated at random intervals at the edge of the buffer region 0.2 µm from the nuclear surface, oriented radially, and polarized toward the cell membrane.

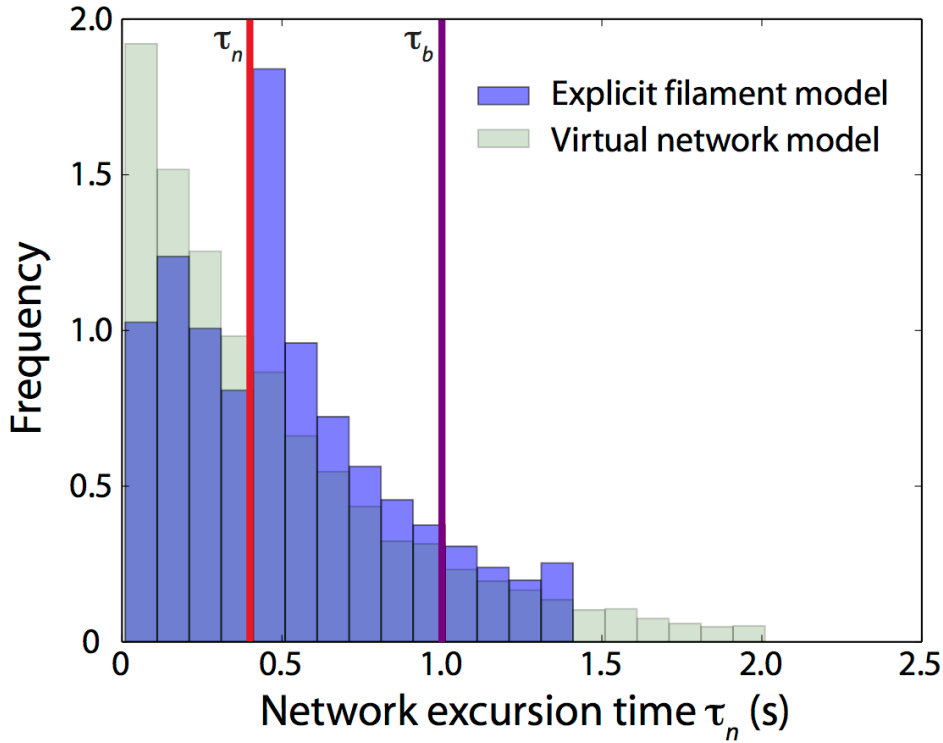

**Figure S5: Simulations of transport along explicit filament networks have a narrower distribution of network excursions.** Explicit filament networks consisted of 450 randomly distributed filaments of length 3 μm. We also measured the mean values of the time intervals during which cargos interacted with filaments ($\tau_n$) and intervals spent in the bulk ($\tau_b$). Extracted $\tau_n$ and $\tau_b$ values were then used to create a virtual filament model of transport that used an exponential distribution of network and bulk residence times with the same average values of $\tau_n$ and $\tau_b$, respectively. Time intervals during which cargos interacted with filaments ($\tau_n$) have a mean value of 0.43 s, and intervals spent in the bulk ($\tau_b$) have a mean of 1.0 s. In comparison to a virtual network model constructed with identical values of $\tau_n$ and $\tau_b$, virtual network simulations have an exponential distribution of $\tau_n$ (by construction) while explicit network simulations have a distribution that deviates substantially. Mean $\tau_n$ (red) and $\tau_b$ (purple) for both distributions are shown as vertical lines.

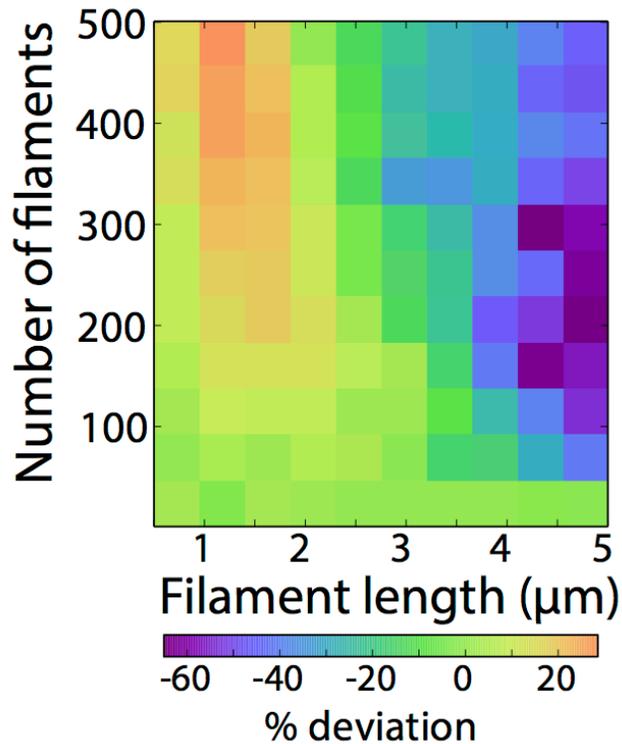

Figure S6: **MFPTs from simulations of transport along explicit filaments can differ significantly from virtual network model simulations.** Explicit filament network simulations were performed at each set of parameters to determine the MFPT. We also measured the mean values of the time intervals during which cargos interacted with filaments ($\tau_n$) and intervals spent in the bulk ($\tau_b$). Extracted $\tau_n$ and $\tau_b$ values were then used to create a virtual filament model of transport that used an exponential distribution of network and bulk residence times with the same average values of $\tau_n$ and $\tau_b$, respectively. The percent deviation plotted is the deviation between the MFPTs of the virtual model and the explicit filament model at equal $\tau_n$ and $\tau_b$.

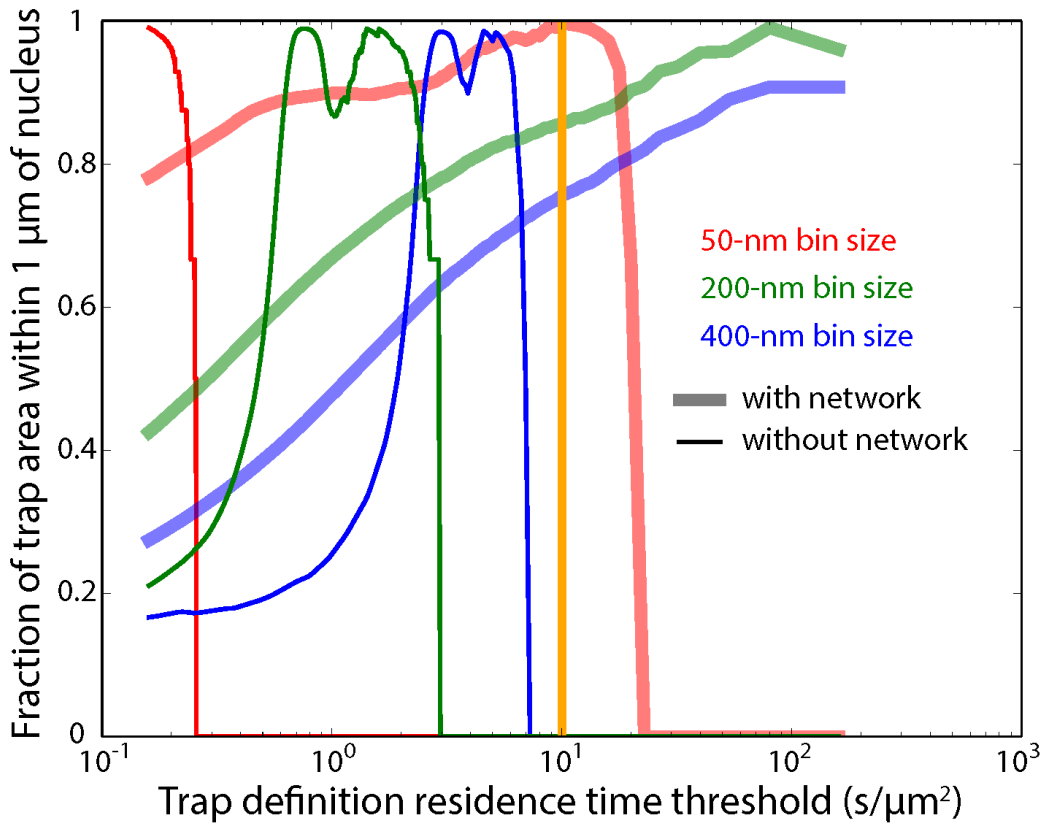

**Figure S7: Cargo trapping as a function of network architecture is insensitive to the binning size used to define traps.** Simulations of 10,000 cargos were performed over 1,000 realizations of networks with 150 filaments of length 3 µm, and also in the absence of a network. Analyses were performed using binning sizes of 50, 200, and 400 nm, over a wide range of residence time thresholds. For a residence time threshold of 10 s/µm$^2$ (orange vertical line), there are no traps in the absence of a filament network, while there is a large fraction of traps located with 1 µm of the nuclear surface in the presence of a filament network for all three binning sizes. For reference, the trap fraction within 1 µm of the nuclear surface, if traps were uniformly distributed throughout the cell, would be only 0.147, as calculated from the ratio of the area of the 1-µm band around the nucleus to the area of the cytoplasm.

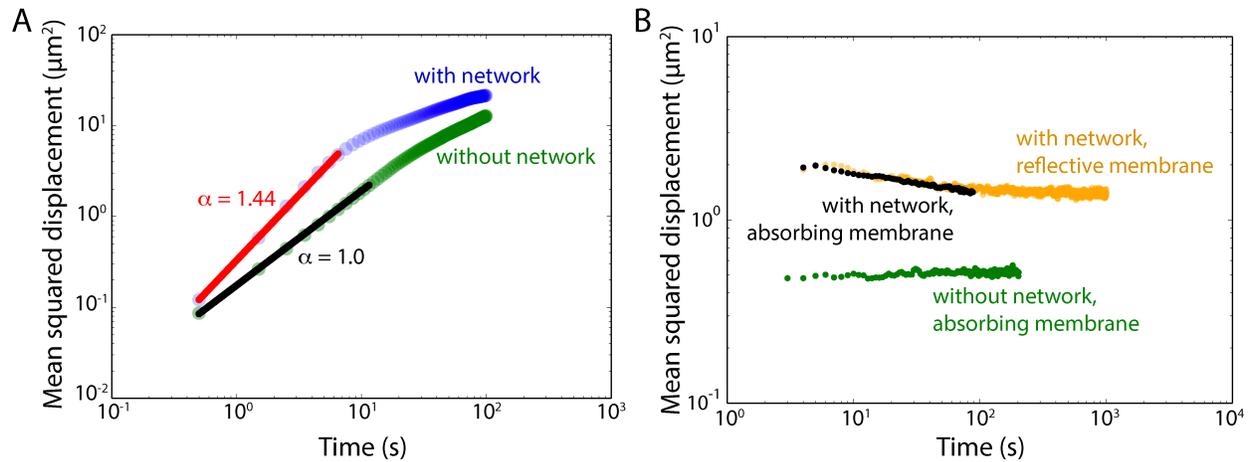

**Figure S8: Ergodicity and lack of aging for transport along explicit filament networks.** A) Time-averaged mean squared displacement (MSD) for explicit filament network simulations of 300 cargos averaged over 100 realizations of networks consisting of 150 filaments of length 3 µm. Over short lag times (up to ~8 s), cargo motion in our explicit network simulations was superdiffusive, with exponent α = 1.44 (where MSD $\sim t^\alpha$). When no network is present, α = 1, reflecting the pure diffusive motion of the cargo over short timescales, before the cell geometry became confining over longer times. B) Ensemble-averaged MSD as a function of cargo trajectory time, with a fixed lag time of ~3 s, for simulation parameters identical to those in (A) and explicit networks that have either an absorbing (black) or reflective cell membrane (yellow). Mean MSD is also shown for trajectories of cargos transported through cells that have no network and an absorbing cell membrane (green). MSD is plotted for trajectory times up to the average cargo trajectory duration. Simulations with a reflective cell membrane were performed to allow for extended timescales of measurement (up to 1000 s), as simulations with an absorbing membrane end quickly. The MSDs converge at ~100 s, indicating that aging is not present and the system is ergodic.